\documentclass[preprint2]{aastex}
\begin{document}

\title{An Imaging Fabry-P\'{e}rot system for the Robert Stobie Spectrograph on the Southern African Large Telescope}
\author{Naseem Rangwala\footnote{email: rangwala@physics.rutgers.edu} and T. B. Williams}
\affil{Department of Physics and Astronomy, Rutgers University, 136 Frelinghuysen Road,
Piscataway, NJ 08854}
\author{Chris Pietraszewski}
\affil{IC Optical Systems Ltd, 190-192 Ravenscroft Road, Beckenham, Kent, BR3 4TW, UK}\and
\author{Charles L. Joseph}
\affil{Department of Physics and Astronomy, Rutgers University, 136 Frelinghuysen Road,
Piscataway, NJ 08854}

\begin{abstract}
We present the design of the Fabry-P\'{e}rot system of the Robert Stobie Spectrograph on the 10-meter class
Southern African Large Telescope and its characterization as measured in the laboratory. This
system provides spectroscopic imaging at any desired wavelength spanning a bandpass 430 -- 860~nm,
at four different spectral resolving powers ($\lambda/\Delta\lambda$) ranging from 300 to 9000.
Our laboratory tests revealed a wavelength dependence of the etalon gap and parallelism with a
maximum variation between 600 -- 720~nm that arises because of the complex structure of the
broadband multi-layer dielectric coatings. We also report an unanticipated optical effect of this
multi-layer coating structure that produces a significant, and wavelength dependent, change in the
apparent shape of the etalon plates. This change is caused by two effects: the physical
non-uniformities or thickness variations in the coating layers, and the wavelength dependence of
the phase change upon reflection that can amplify these non-uniformities. We discuss the impact of
these coating effects on the resolving power, finesse, and throughput of the system. This
Fabry-P\'{e}rot system will provide a powerful tool for imaging spectroscopy on one of the world's
largest telescopes.
\end{abstract}


\section{Introduction}
The Southern African Large Telescope (SALT) is a 10-meter class telescope owned and operated by an
international consortium of partners. The telescope design (\cite{stobie}; \cite{arek}) is a
modification of the Hobby-Eberly Telescope, employing a multi mirror spherical primary that is
held fixed during an observation. A prime-focus tracker follows the image and contains a spherical
aberration corrector and instruments. The Robert Stobie Spectrograph (RSS) \cite{eric, Ken, Kob,
mike} is a prime-focus spectrograph providing long-slit, multi-object, and imaging spectroscopic
modes, together with polarimetric capability. The RSS was designed, built and commissioned as a
collaboration between groups from the University of Wisconsin and Rutgers University.

One of the goals of the RSS is to provide imaging spectroscopic capabilities for SALT, using
Fabry-P\'{e}rot (FP) technology. The Rutgers group has extensive experience in this field, having
provided the Rutgers Fabry-P\'{e}rot (RFP) instrument \cite{rfp} at the Cerro Tololo Observatory
for many years, and designing a new FP instrument, ARIES, for the SOAR telescope \cite{aries}. The
RSS FP system represents an evolution of these designs, adapted for the characteristics of the
SALT telescope and the scientific requirements of the SALT partners. The major design goals of the
system are to provide spectroscopic imaging over the full $8\arcmin$ field of view of the
telescope, at a variety of spectral resolutions ranging from 500 to 12500, over the wavelength
range from 430 to 860 nm.

The use of FP technology in astronomy for imaging spectroscopy has evolved over the years and is
now being successfully employed on a regular basis by many groups around the world for galactic
and extragalactic studies. The first use of the FP in astronomy was by Fabry himself
(\cite{fabry}) to study the Orion nebula. De Vaucouleurs developed the ``galaxymeter"
(\cite{galaxymeter}) for measuring galaxy rotation curves. In the 1980s, a series of Taurus FP
systems were developed and used at several large telescopes (\cite{atherton2}). Other systems
followed, including the Ohio state imaging FP spectrometer (\cite{richard}), the Goddard FP imager
(\cite{gelderman}), the Hawaii imaging FP interferometer (\cite{tully}), and the Wisconsin H-alpha
imager (\cite{tufte}). The Taurus tunable filter (\cite{jones}) pioneered the use of FP techniques
for lower resolution studies, followed by the NACO FP tunable filter at VLT (\cite{hartung}) and
the Magellan-Maryland tunable filter. The UNSWRIF system (\cite{ryder}) extends FP technology into
the near infrared.

This paper documents the design of the SALT RSS FP system and its characteristics as measured in
our laboratory. It is organized as follows: in section \ref{basic} we review the basic properties
of the FP interferometer; in section \ref{design} we present the design of the FP subsystem for
the SALT RSS; we present the results of our laboratory testing of the FP etalons in section
\ref{result}; we discuss the unanticipated effects of the coatings in section \ref{coat} and
conclude in section \ref{conc}. In a subsequent paper we plan to describe the commissioning of the
FP subsystem on SALT and document its performance characteristics in the observing environment.

\section{Basic Principles and Characteristics of a Fabry-P\'{e}rot
Interferometer}\label{basic}
\subsection{Ideal Fabry-P\'{e}rot}
A FP etalon consists of a pair of parallel glass plates and works on the principle of multiple
beam interference. High reflectivity coatings are applied to the inner surfaces of the plates, and
anti-reflection coatings are applied to the outer surfaces. Light entering the etalon at an angle
$\theta$ will undergo multiple reflections between the plates, resulting in an interference
pattern, subject to the following condition:

\begin{equation}\label{intcond}
2d\cos\theta = N\lambda
\end{equation}

\noindent where $d$ is the gap between the plates, $\theta$ is the incident angle relative to the
plate normal, $N$ is the order of interference, and $\lambda$ is the wavelength of the light in
the gap medium ($\lambda = \lambda_{0}/\mu$, where $\lambda_{0}$ is the wavelength in vacuum and
$\mu$ is the refractive index). The separation between any two consecutive orders of interference
is called the free spectral range (FSR), which is related to the gap by
\begin{equation}\label{efsr}
FSR = \frac{\lambda^2}{2d}
\end{equation}

\begin{figure*}[t]
  \includegraphics[scale=0.50]{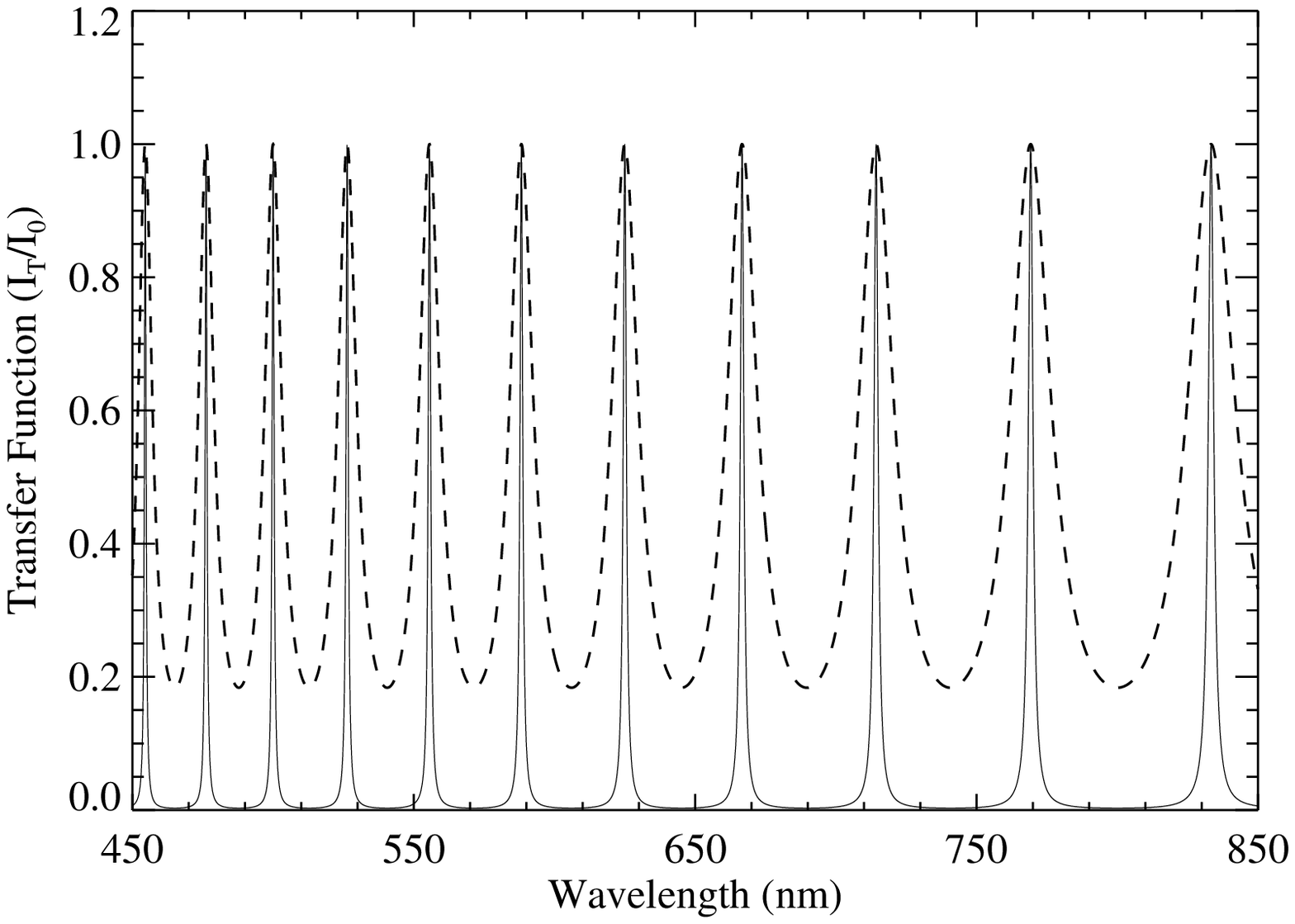}
  \includegraphics[scale=0.40]{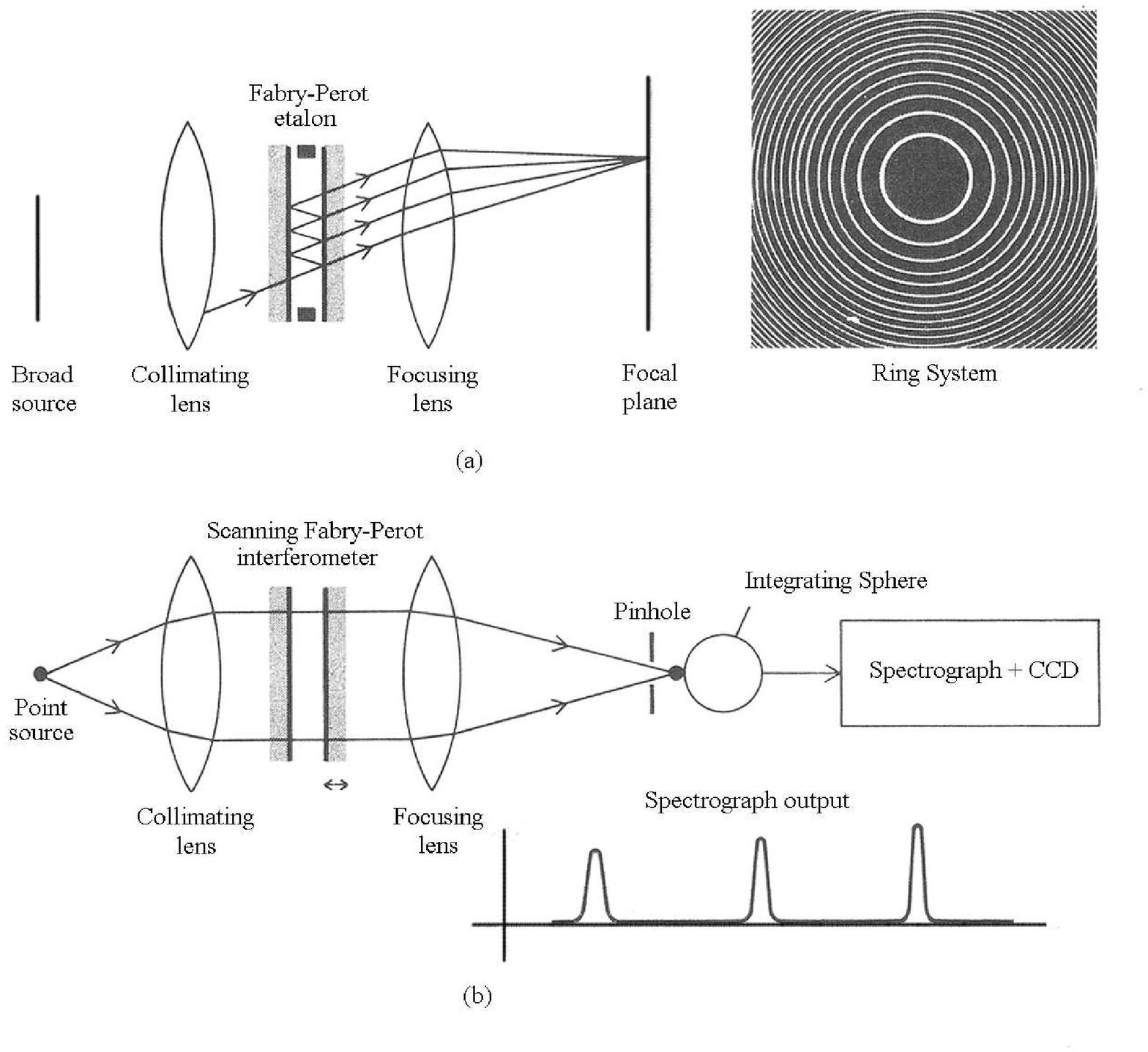}
  \caption{\label{airy}Left: The Airy function (eg.~$\ref{eairy}$)
transmission profiles of ideal Fabry-P\'{e}rot etalons with $d = 5\mu$m and $R = 0.8$ (solid) $\&$
$0.45$ (dashed). Note that the FSR increases as wavelength increases. Right: Figure taken from
\cite{fowles} (used with permission): (a) an extended source of light produces a concentric ring pattern and (b)
laboratory setup for measuring etalon properties.}
\end{figure*}

The transmission function of an ideal FP is an Airy function (left panel of Figure $\ref{airy}$)
given by

\begin{equation}\label{eairy}
I_{T} = I_{0}\frac{T^{2}}{(1 - R)^{2}} \frac{1}{1 +
F\textrm{sin}^{2}(\Delta/2)}
\end{equation}

\noindent where $I_{0}$ is the intensity of the incident beam, $R$
and $T$ are the reflectance and transmittance of the high
reflectance coatings, $\Delta = 4\pi d \textrm{cos}\theta/\lambda$
is the phase difference between two successive reflections, and $F =
4R/(1-R)^{2}$. (Note that ideal coatings will have no absorption or
scattering, so $T = 1 - R$ and the peak transmission of the etalon
is 100\%.)

Finesse is a measure of sharpness of the interference fringes and is measured experimentally by
the ratio of the FSR to the full width at half maximum (FWHM) of the transmission profile. For an
ideal etalon, the finesse is determined solely by the reflectance $R$ and is termed the
``reflectance finesse", $\mathcal{F}_{R}$:
\begin{equation}\label{rfin}
\mathcal{F}_{R} = \frac{\pi}{2} \sqrt{F} = \pi\frac{\sqrt{R}}{(1-R)}
\end{equation}
The spectral resolving power of the FP is given by
\begin{equation}\label{rp}
\mathcal{R} \equiv \frac{\lambda}{\delta\lambda} = N\mathcal{F}_{R}
\end{equation}
Therefore the resolving power can be increased by either increasing
the order of the interference ($N \propto 2d/\lambda$) or by
increasing the reflectance of the coatings.

In the case of an extended monochromatic light source the interference pattern takes the form of a
ring of radius $r$, with $\lambda \propto 1/r^{2}$. If the illumination subtends a sufficiently
large range of angles $\theta$, then multiple concentric rings of successive orders are formed
(right panel of Figure~$\ref{airy}$(a)). The transmitted wavelength can be selected by either
changing the gap spacing (mechanical scanning) or by changing the index of the medium in the gap
(pressure scanning). In this way the FP etalon acts as a tunable narrow band filter.
\subsection{Practical Fabry-P\'{e}rot}
Practical high-reflectance coatings will have some non-zero absorption coefficient $A$, making the
transmission $T = 1 - R - A$. The expression for the etalon transmission is still given by
equation~$\ref{eairy}$, but the maximum transmission is no longer 100\%. The effect of absorption
is to lower the transmission of the fringes, but does not affect the finesse or spectral
resolution of the etalon. To minimize the absorption $A$, most current astronomical FP systems
employ multi-layer dielectric coatings. A typical value of $A$ for such coatings is 0.005; for $R
= 0.90$, the maximum transmission is reduced to 90\%. It is usually desirable for the coatings to
have as broad a spectral bandwidth as possible; current designs can achieve an octave in
bandwidth, but typically degrade rapidly at longer or shorter wavelengths.

While the plates of the FP interferometer are highly polished, they are not perfectly flat. This
non-flatness gives rise to a defect finesse ($\mathcal{F}_{D}$). The effective finesse,
$\mathcal{F}$, of the etalon is then the combination of the reflectance finesse and the defect
finesse given by
\begin{equation}\label{tfin}
\frac{1}{\mathcal{F}^{2}} = \frac{1}{\mathcal{F}_{R}^{2}} +
\frac{1}{\mathcal{F}_{D}^{2}}
\end{equation}
Three common types of defects are: (1) curvature in the plates ($\mathcal{F}_{Dc}$), (2) surface
irregularities or plate roughness ($\mathcal{F}_{Dr}$) and (3) departure from parallelism
($\mathcal{F}_{Dp}$). \cite{atherton} give the following expressions for these defects:
\begin{eqnarray}
\label{dfin}
\mathcal{F}_{Dc}&=&\frac{\lambda}{2\delta t_{c}}\\
\mathcal{F}_{Dr}&=&\frac{\lambda}{4.7\delta t_{r}}\\
\mathcal{F}_{Dp}&=&\frac{\lambda}{\sqrt{3}\delta t_{p}}
\end{eqnarray}
where $\delta t_{c}$, $\delta t_{r}$, $\delta t_{p}$ are measures of
the deviations for their respective defects.
The expression for $\mathcal{F}_{Dc}$ assumes spherical curvature;
in general, the curvature can have a complicated shape.
The total defect finesse is the combination of all the above defects
\begin{equation}
\frac{1}{\mathcal{F}_{D}^{2}} = \frac{1}{\mathcal{F}_{Dc}^{2}} +
\frac{1}{\mathcal{F}_{Dr}^{2}} + \frac{1}{\mathcal{F}_{Dp}^{2}}
\end{equation}
In addition to physical defects, optical effects in multi-layered reflection coatings can produce
additional \textit{apparent} distortions of the plates. This effect is discussed in detail in
later sections. Any departure from flatness, either physical or apparent, will lower the overall
finesse, spectral resolution and transmitted intensity of the etalon.

\section{FP System for SALT}\label{design}
The design goals for the SALT FP system are to provide spectroscopic imaging over a wide range of
wavelengths and spectral resolutions, while fitting within the size, weight, and budget
constraints of the RSS. A survey of the SALT user community showed there was interest in programs
at wavelengths extending from the ultraviolet atmospheric limit to the near-infrared. We could not
accommodate this wide spectral range with a single set of etalon coatings, and chose a bandwidth
of $430 - 860$~nm, leaving the range of shorter wavelengths for a possible future set of etalons.
The SALT user survey also showed interest in a wide range of spectral resolutions, so we decided
on a system employing three etalons, with gap spacings of 5, 27, and 135~$\mu$m, providing
spectral resolutions at 650~nm of 500, 2500, and 12500, respectively. We shall subsequently refer
to these etalons as SG, MG, and LG (for small, medium and large gap). The diameter of the
collimated beam in the RSS is 150 mm, so etalons with this clear aperture are required. It is
difficult to obtain etalon plates of this size with surface quality better than $\lambda / 100$,
so $\mathcal{F}_{D}\sim 50$. We chose a reflectance $R = 0.90$ for the coatings, giving
$\mathcal{F}_{R} = 30$, and the combined effective finesse $\mathcal{F}\sim 25$.

\begin{figure*}
  \epsscale{2.0}
  \plotone{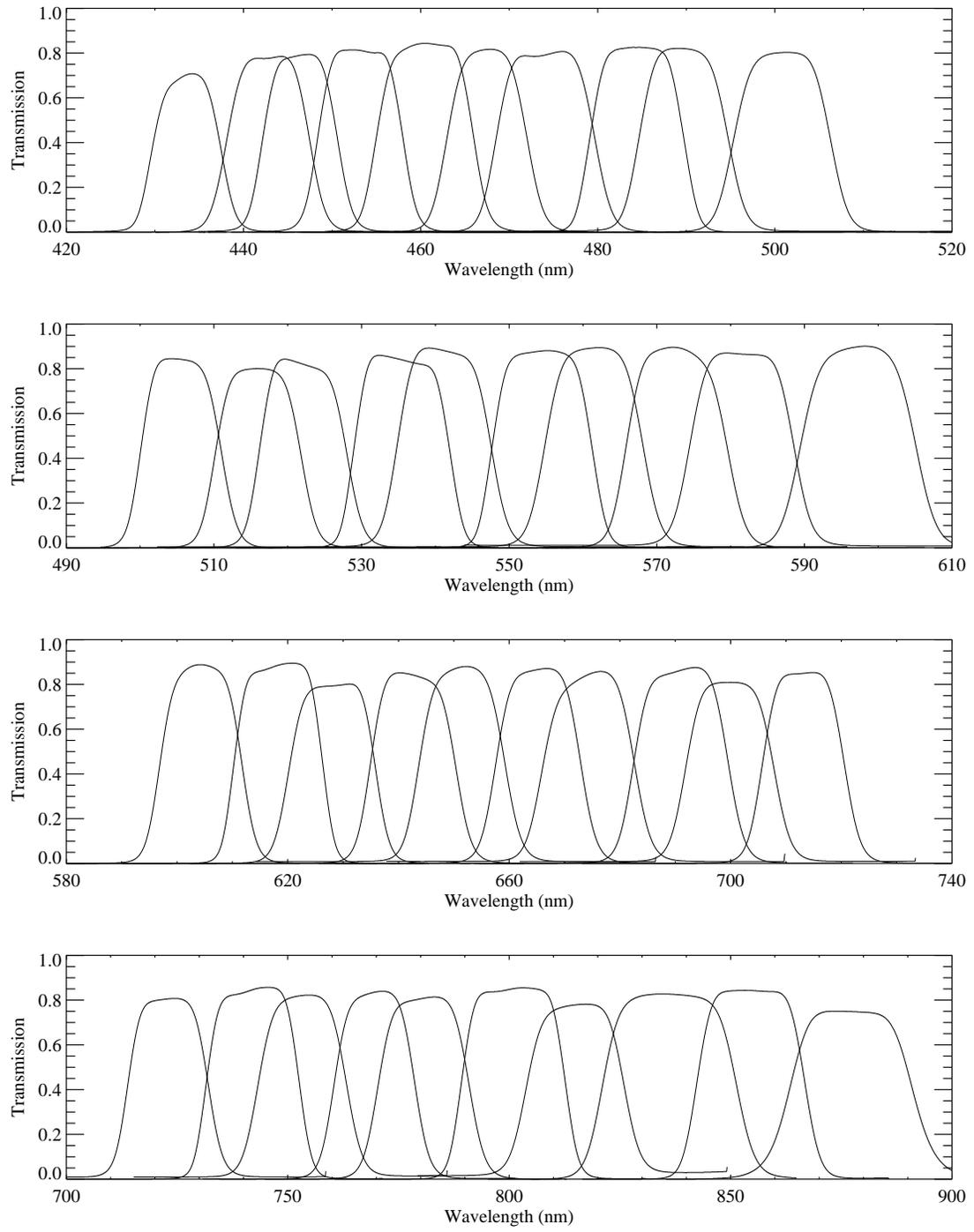}\\
 \caption{Transmission curves of the order selecting interference filter set}\label{filter}
\end{figure*}

A blocking filter is required to select the desired transmission order of the FP etalon. For the
low resolution etalon SG, with larger FSR, a reasonable number of standard interference filters
can be used to isolate the orders. We designed a set of filters to isolate each of the orders of
the SG etalon within our wavelength range. The filters are 4-period designs to give relatively
flat-topped transmission profiles with steep sides. The design goal is to choose filter central
wavelengths and widths so that in the worst case, when the etalon is tuned to the cross-over
wavelength between adjacent filters, the filter adequately suppresses the transmission of the
neighboring order of the etalon. The filter set consists of 40 filters, whose measured
transmission curves are shown in Figure~\ref{filter}.

The transmissions are typically 80\% or better, less at the crossover wavelengths between adjacent
filters. Since the etalon FSR increases with wavelength, the filters are narrower and more closely
spaced at the short wavelength end of the range. Manufacturing tolerances have produced some gaps
in the set where the transmission at the crossover wavelength is unacceptably low; additional
filters will be purchased in the future to fill these gaps.

\begin{figure*}
  \includegraphics[scale=0.65,angle=90]{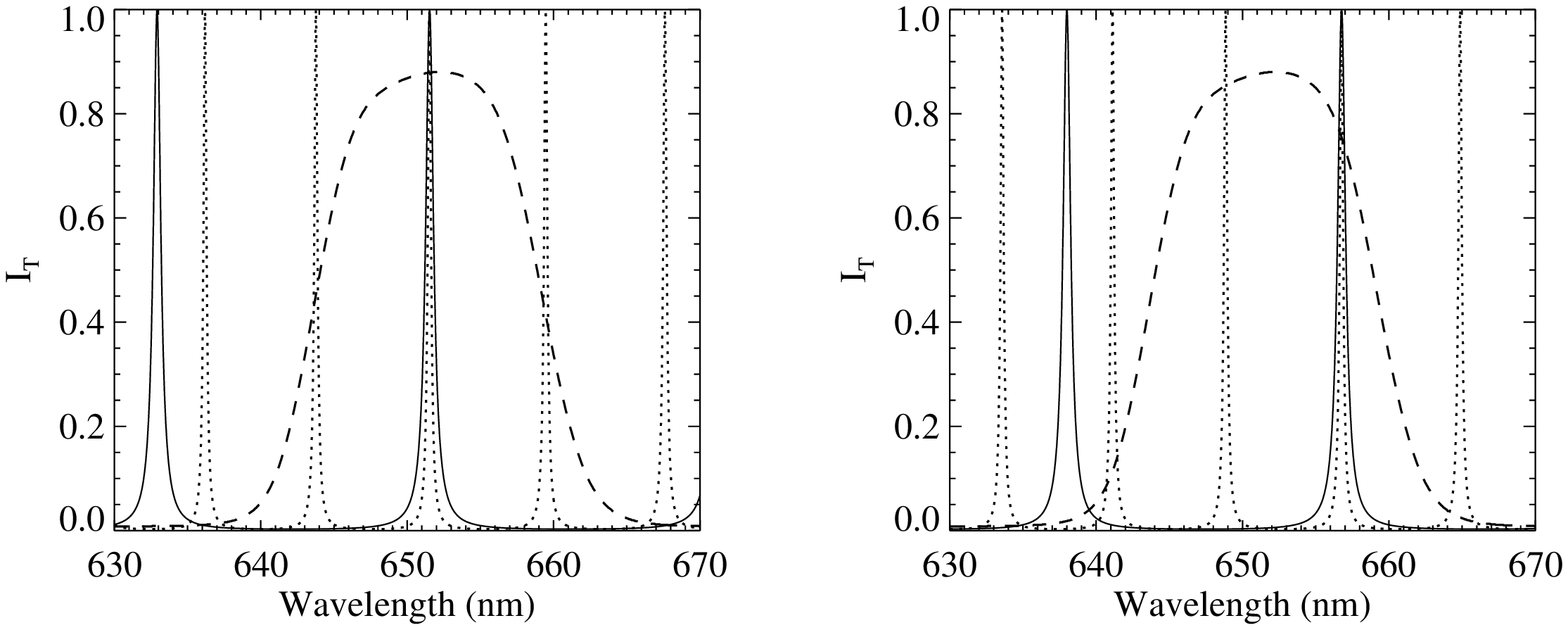}\\
 \caption{Order selection with interference filter and dual etalons.
  Solid curve: SG etalon; dashed curve: filter; dotted curve: MG etalon.}\label{dual}
\end{figure*}

An example of the filter blocking for the SG etalon is shown in Figure~$\ref{dual}$, where the
solid curve is the transmission of an ideal etalon of gap 11~$\mu$m and the dashed curve is the
measured transmission of one of the filters. The left panel shows the etalon tuned to the central
wavelength of the filter, and the adjacent orders are highly blocked. The right panel shows the
worst case when the etalon is tuned to the crossover wavelength to the next filter, and the
``parasitic light" leakage of the adjacent order is a few percent.
For the higher resolution etalons with their smaller FSR, the blocking scheme discussed above
becomes impractical, requiring a very large number of narrow (hence expensive and lower peak
transmission) filters. We thus decided to employ a dual etalon system for the higher resolution
modes. Either the MG or LG etalon is used in series with the SG etalon and one of the blocking
filters discussed above. The SG etalon is tuned to the same wavelength as the higher resolution
etalon, and suppresses its adjacent orders. As above, the filter suppresses the adjacent orders of
the SG etalon. The scheme is illustrated in Figure~$\ref{dual}$, where an idealized MG etalon
(27~$\mu$m gap) transmission is shown by the dotted curve. The total transmission (not shown in
the figure for clarity) is the product of the three curves. Parasitic light is highly suppressed
in the medium resolution mode. In the highest resolution mode, the wings of the SG transmission do
not fully block the adjacent orders of the LG etalon, producing parasitic light of typically 5\%.
The dual etalon design makes it possible to select any desired wavelength within the coating
bandpass, and at any of the etalon resolutions, with a limited number (40) of interference
filters.

We selected IC Optical Systems Ltd (ICOS) to supply the etalons and their controllers. Each etalon
is controlled by a three-channel system that uses differential capacitance micrometers and
piezo-electric actuators, incorporated into the etalon, to monitor and correct errors in plate
parallelism and spacing. Two channels (X and Y) control the parallelism using capacitor pairs
optically contacted to the etalon plates. The third channel (Z) sets and maintains the spacing
between the plates by comparing another capacitor optically contacted to the plates to a fixed-gap
reference capacitor. Since all the measurements are differences between air-gap capacitors within
the ambient etalon environment, they are highly stable and relatively immune to the influence of
changing temperature, humidity, etc. The system can move the plates over a range of $\sim$
6~$\mu$m with a resolution of 0.49~nm.

The range of the piezo-electric actuators is sufficient to change the gap of the SG etalon by a
factor of two, allowing this etalon to be operated in two different spectral resolution modes.
Thus the system provides four distinct spectral resolutions. Setting the SG etalon to its smallest
gap ($\sim$5~$\mu$m) produces the Tunable Filter (TF) mode with $\mathcal{R} \sim350$, and setting
it to its largest gap ($\sim$11~$\mu$m) produces the Low Resolution (LR) mode with $\mathcal{R}
\sim750$. For the larger gap etalons, the piezo actuators cannot change the gap enough to produce
significantly different resolutions, and we choose to operate the etalons near their largest gap
settings for maximum resolution. Setting MG etalon to its largest gap ($\sim$28~$\mu$m) produces
Medium Resolution (MR) mode, with $\mathcal{R} \sim1500$. Setting LG to its largest gap
($\sim$136~$\mu$m) produces High Resolution (HR) mode, with $\mathcal{R} \sim8500$. Details on
these modes are presented in section \ref{result}.
\begin{figure}[t]
  \includegraphics[scale=0.45]{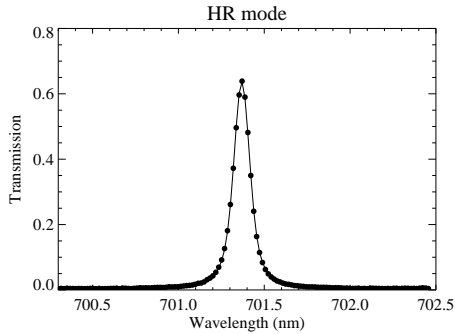}
  \caption{Transmission profile of the LG etalon is well represented by
   a Voigt function (solid line)}
  \label{voigt}
\end{figure}
\section{Lab Testing}\label{result}
The characterization and calibration of the FP system was carried out in our lab at Rutgers
University. The test setup, shown schematically in right panel of Fig.~\ref{airy}(b), consisted of
a pair of telescopes, each with 1200 mm focal length and 150 mm aperture, used as a collimator and
a camera. The etalons were mounted on a slide that allowed them to be positioned into and removed
from the collimated beam. The light source used was a 100~W quartz halogen lamp. The transmitted
light was fed into an integrating sphere that illuminated the entrance slit of a 0.5m spectrograph
with a CCD detector. The spectrograph allows to measure the transmission profile of the FP over
several orders in a single spectrogram. A typical output transmission profile as shown in Figure
\ref{voigt} was fit very well by a Voigt function, providing the central wavelength and FWHM of
each individual order. The spectrograph used three gratings that provided spectral resolutions of
0.25~nm, 0.08~nm, and 0.04~nm. The TF and LR modes, with 13 and 26 orders respectively, were
measured with the 0.25~nm resolution. The MR mode with 72 orders was measured with a combination
of the 0.25~nm and 0.08~nm resolution. The HR mode with 330 orders was measured with a combination
of the 0.08~nm and 0.04~nm resolution. The measured characteristics of all modes and orders are
listed in Table \ref{main}.

\begin{figure*}
  \includegraphics[scale=0.65,angle=90]{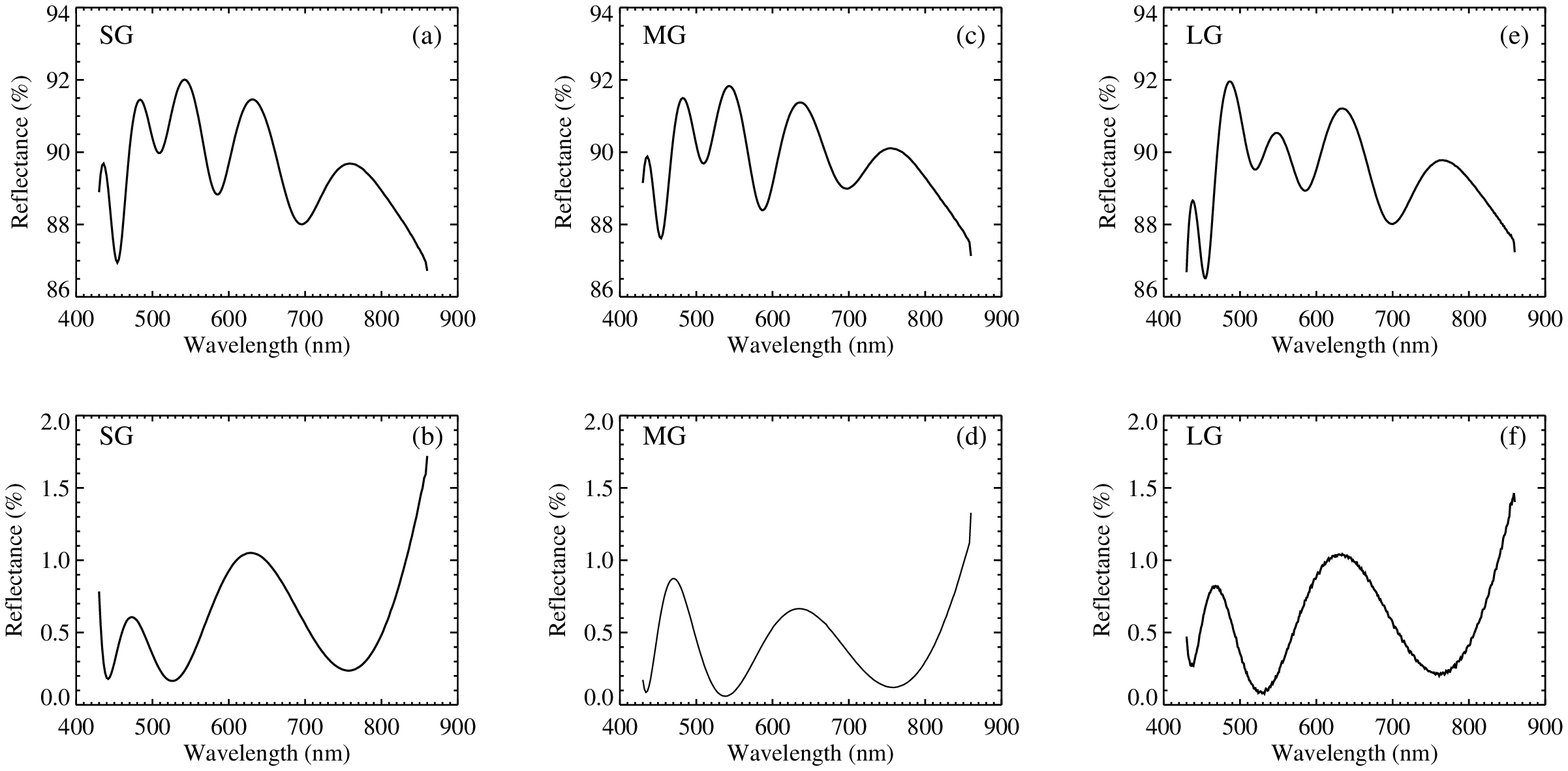}\\
  \caption{High reflection and anti-reflection coating curves of
   the etalon plates, provided by ICOS.}
  \label{coating}
\end{figure*}

We do not have the facilities in our lab to measure directly the properties of the etalon
coatings. Figure~$\ref{coating}$ shows the measurements supplied by ICOS of the high reflection
coatings at the etalon gaps and the anti-reflection coatings on the outer surfaces for each of the
three etalons. The coatings meet our specifications of $90 \pm 4\%$ and $<1\%$, respectively.
\subsection{Results}
\begin{deluxetable}{ccccccccccc}
\tablecaption{Properties of the TF, LR, MR and HR modes.} \tablehead{\colhead{wavelength}&
\colhead{FWHM} & \colhead{FSR} & \colhead{finesse} & \colhead{resolution} & \colhead{gap (d)} &
\colhead{X offset} & \colhead{Y offset} \\
\colhead{(nm)} & \colhead{(nm)}& \colhead{(nm)} & \colhead{} & \colhead{} & \colhead{($\mu$m)} &
\colhead{} & \colhead{}} \startdata  \multicolumn{8}{c}{TF Mode}\\ \hline
 440.875   &   1.595   &   23.270   &   14.60   &      276   &      4.18   &   -112   &     40\\
 464.144   &   1.823   &   24.810   &   13.60   &      255   &      4.34   &   -100   &     25\\
 490.500   &   1.663   &   28.040   &   16.90   &      295   &      4.29   &   -112   &     33\\
 520.218   &   1.888   &   31.160   &   16.50   &      276   &      4.34   &   -112   &     33\\
 552.812   &   1.912   &   34.410   &   18.00   &      289   &      4.44   &   -120   &     40\\
 589.043   &   2.353   &   37.440   &   15.90   &      250   &      4.63   &   -112   &     45\\
 627.701   &   1.973   &   38.380   &   19.50   &      318   &      5.13   &   -165   &     85\\
 665.800   &   2.033   &   35.040   &   17.20   &      327   &      6.33   &   -250   &    140\\
 697.783   &   1.874   &   32.830   &   17.50   &      372   &      7.42   &   -290   &    180\\
 731.461   &   1.847   &   36.280   &   19.60   &      396   &      7.37   &   -280   &    160\\
 770.346   &   2.093   &   42.060   &   20.10   &      368   &      7.05   &   -245   &    140\\
 815.590   &   2.514   &   47.930   &   19.10   &      324   &      6.94   &   -255   &    150\\
 866.214   &   3.093   &   50.620   &   16.40   &      280   &      7.41   &   -220   &    150\\
 \hline
 \multicolumn{8}{c}{LR Mode}\\
 \hline
 426.480   &   1.173   &    8.930   &    7.60   &      364   &     10.18   &    -30   &    -85\\
 435.412   &   0.621   &    9.310   &   15.00   &      701   &     10.18   &    -30   &    -85\\
 444.605   &   0.660   &   10.000   &   15.20   &      674   &      9.88   &      0   &   -110\\
 454.946   &   0.703   &   10.650   &   15.10   &      647   &      9.72   &      0   &   -110\\
 465.886   &   0.699   &   11.220   &   16.10   &      667   &      9.67   &    -10   &   -110\\
 477.368   &   0.612   &   11.760   &   19.20   &      780   &      9.69   &    -10   &   -110\\
 489.406   &   0.630   &   12.360   &   19.60   &      777   &      9.69   &    -10   &   -110\\
 502.079   &   0.677   &   13.010   &   19.20   &      742   &      9.69   &    -10   &   -110\\
 515.371   &   0.710   &   13.690   &   19.30   &      726   &      9.70   &    -10   &   -110\\
 529.743   &   0.706   &   14.380   &   20.40   &      750   &      9.76   &    -20   &    -95\\
 544.213   &   0.698   &   15.080   &   21.60   &      780   &      9.82   &    -10   &   -105\\
 559.896   &   0.751   &   15.840   &   21.10   &      746   &      9.90   &    -20   &    -95\\

\enddata
\tablecomments{Table \ref{main} is published in its entirety in the electronic edition of the {\it
Astronomical Journal}.  A portion is shown here for guidance regarding its form and
content.\label{main}}
\end{deluxetable}
\subsubsection{Parallelism, FSR and Effective Gap}\label{gap}
\begin{figure*}[t]
  \includegraphics[scale=0.4,angle=90]{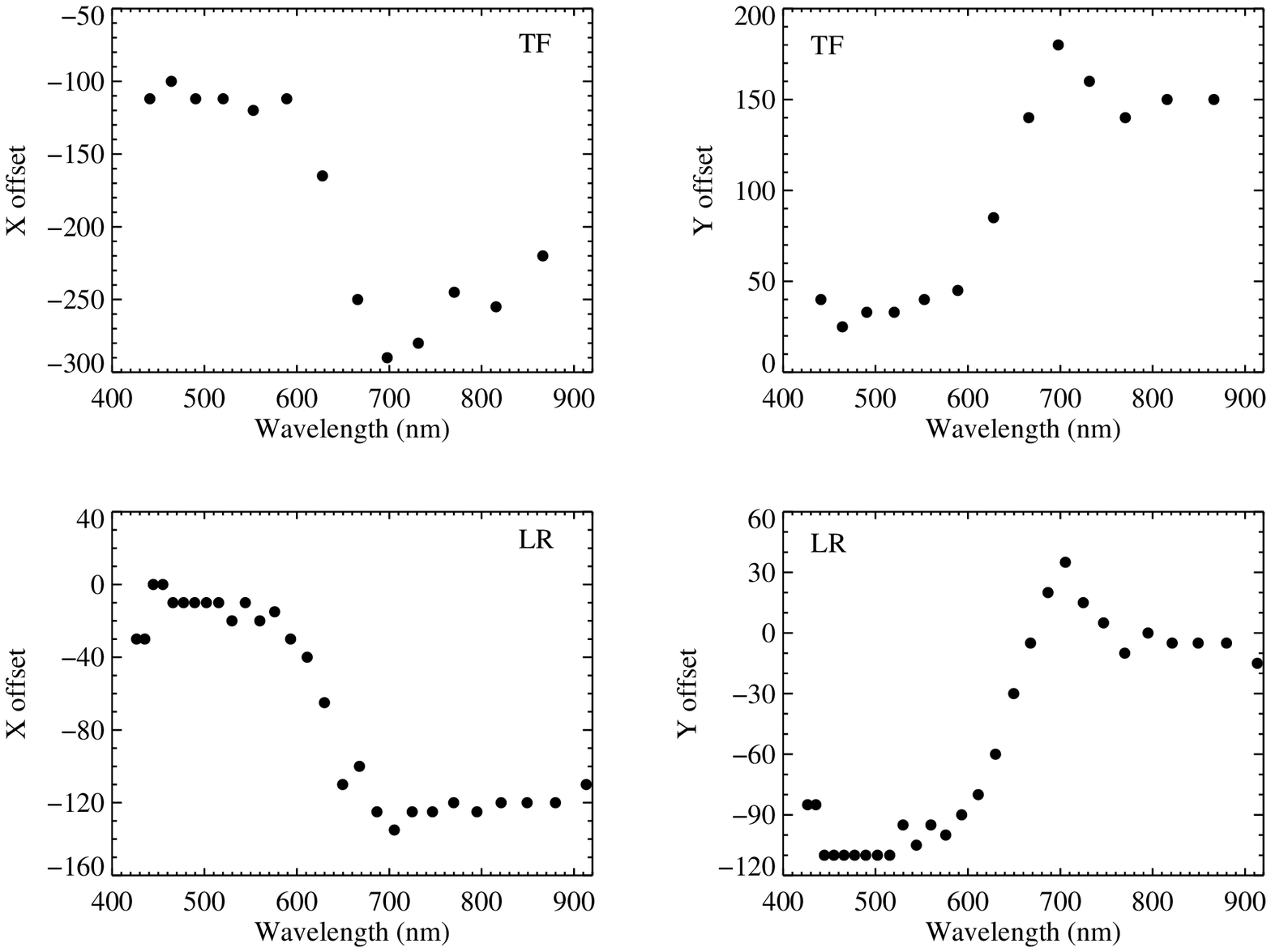}
  \includegraphics[scale=0.4,angle=90]{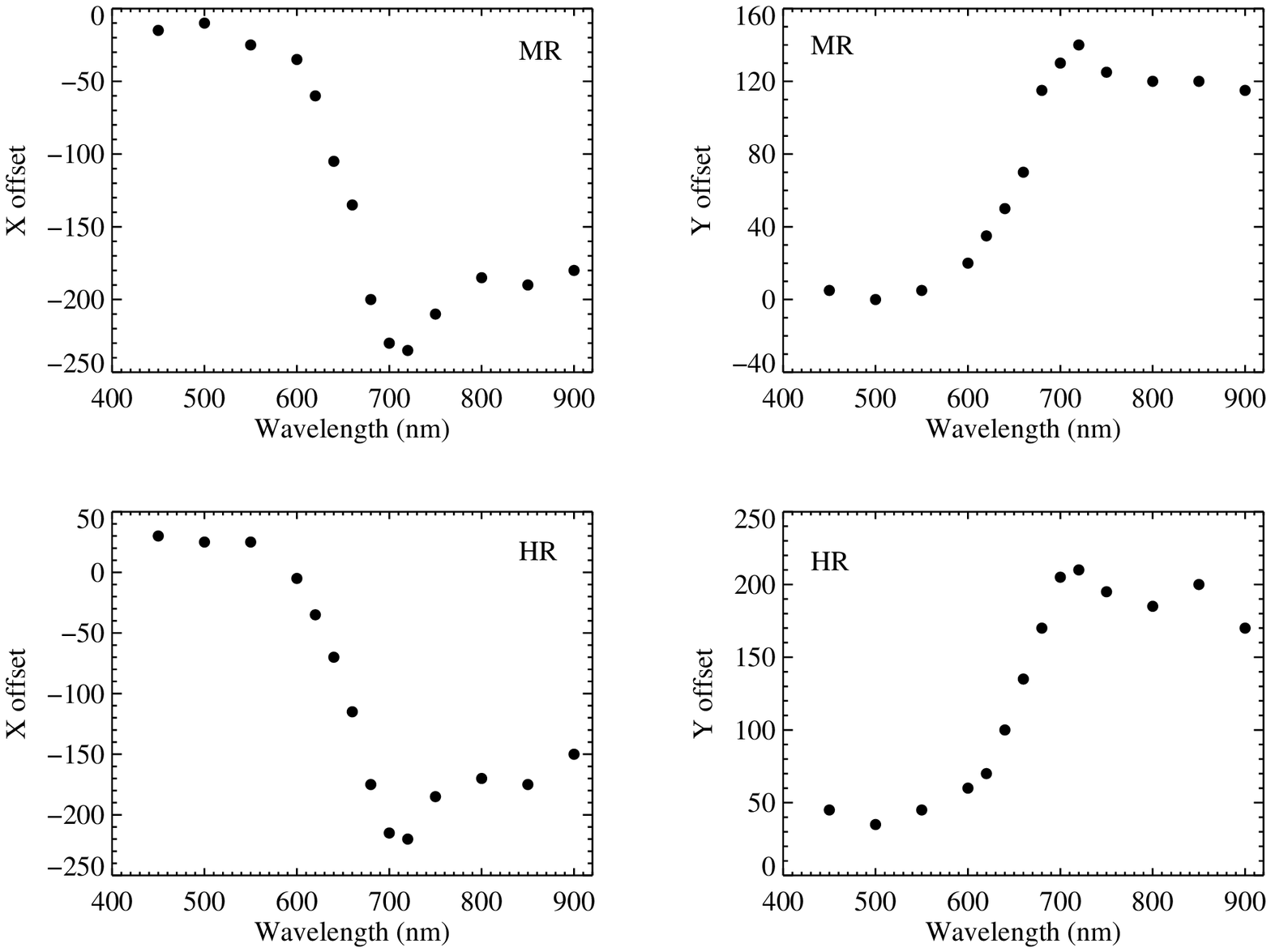}
  \caption{$X$ and $Y$ parallelism curves for the four spectral modes show that the
  parallelism settings are wavelength dependent with maximum variation between 600~nm and
  720~nm.}
  \label{para}
\end{figure*}
The two plates of the etalon should be as parallel to each other as possible in order to achieve
maximum resolution and transmission. The parallelism is adjusted by changing the X and Y offsets
of the controller, the optimal values of which are obtained when the transmission profiles are
most symmetric and sharpest\footnote{The control system settings can be affected by environmental
conditions like temperature and humidity.  More testing will be carried out in the telescope
environment during the commissioning of the instrument to check its stability.}. These values of X
and Y are listed in Table \ref{main} and plotted in Figure $\ref{para}$. The precision of our
determination of the optimum parallelism setting is $\pm 10$ units. For an ideal etalon, the
parallelism settings will be wavelength independent. However, we found that in all four modes the
parallelism settings are constant at shorter wavelengths, vary between 600 -- 750 nm, and are
again constant (but at different values) at longer wavelengths. As seen in Figure~$\ref{para}$,
the shapes of the parallelism setting curves are remarkably similar for all three etalons and all
four resolution modes.

\begin{figure*}[t]
  \includegraphics[scale=0.45]{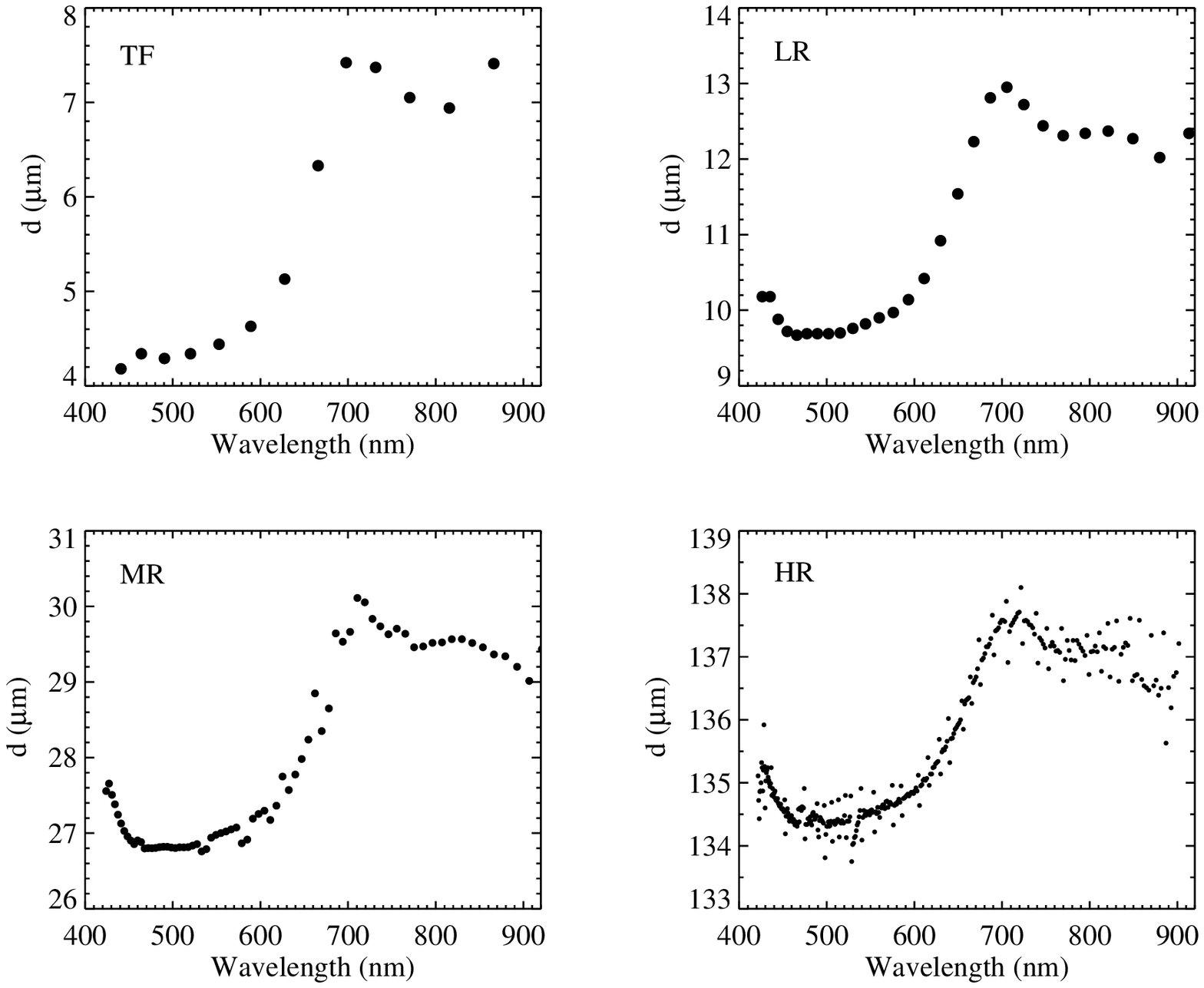}
  \includegraphics[scale=0.45]{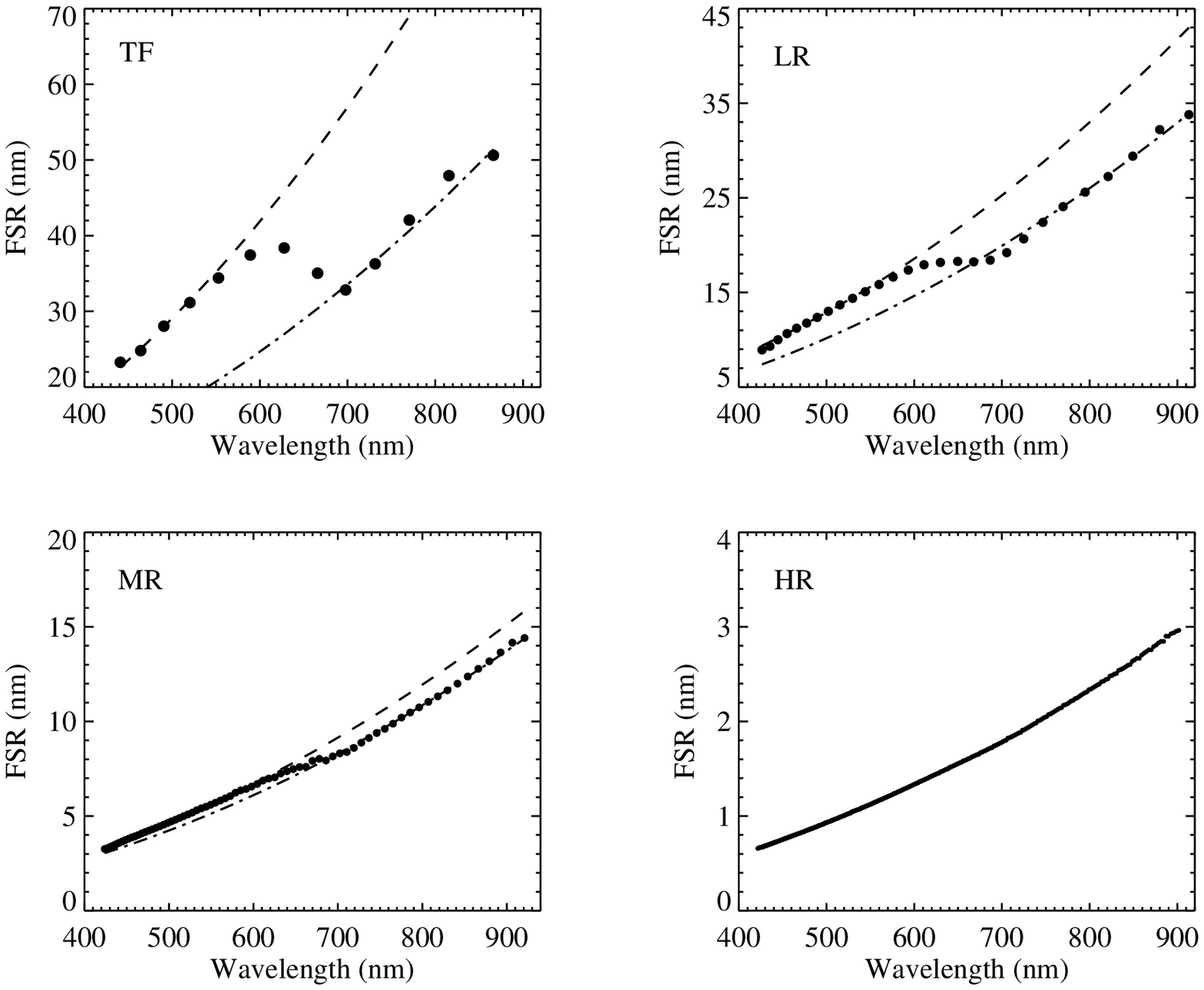}
  \caption{Left: The effective gap also shows wavelength dependence
with maximum variation between 600~nm and 720~nm. Right: FSR changes quadratically with wavelength
for a fixed $d$. The FSR curve shows inflection between 600~nm and 720~nm because the effective
gap changes.}\label{gapfsr}
\end{figure*}

The effective gap of the etalon can be determined by measuring the FSR between a pair of
consecutive transmission orders, and applying equation $\ref{efsr}$. The results, displayed in the
left panels of Figure $\ref{gapfsr}$, show that in all four modes, the effective gap increases by
about 3~$\mu$m in the same 600 -- 750 nm wavelength range in which the parallelism settings
change. Plotting the FSR as a function of wavelength, shown in the right panels of Figure
$\ref{gapfsr}$, illustrates the same effect. The dashed and dot-dashed curves are plots of
equation $\ref{efsr}$ for the best-fit values of the gap at short and long wavelengths,
respectively. In all four modes at short wavelengths the FSR exhibits the quadratic increase with
wavelength characteristic of a fixed gap. Between 600~nm and 750~nm the curves show inflections.
At longer wavelengths the FSR curve again becomes quadratic, but with a larger fixed gap. Again,
the shape of the change of the effective gap is remarkably similar for all the etalons.
The variation of effective gap and parallelism with wavelength is most likely due to the complex
structure of the high reflective coatings that are required to achieve the broad bandwidth from
430~nm to 860~nm of our specification. Similar wavelength dependent effects were also seen by
\cite{ryder} with 70 mm aperture etalons. This unexpected behavior of the coatings will have effects
on other properties of the etalon as well and is discussed in detail in section $\ref{coat}$.

\subsubsection{Plate Flatness}\label{pltflt}
\begin{figure*}
  \includegraphics[angle=90,scale=0.6]{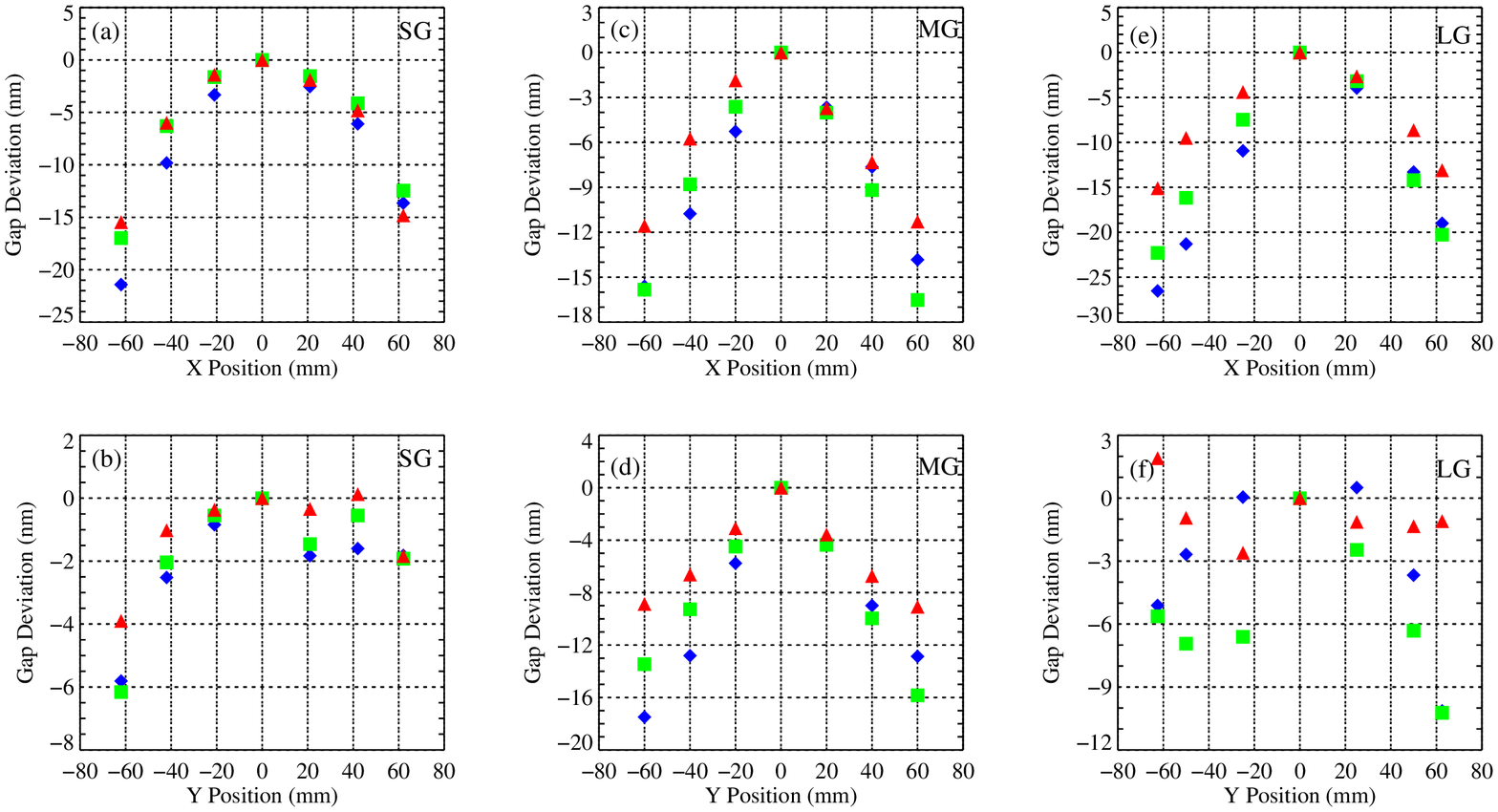}\\
  \caption{Measurements of the effective shape of the etalon plates for
   the three etalons.  The red triangles, green squares and blue diamonds show the
   relative gap at 850, 650 and 500~nm respectively.}\label{flat}
\end{figure*}

As discussed in section 2.2, the finesse and resolution can be affected by deviations from plate
flatness. We measured the effective shape of the plates along the X and Y axis of the etalon by
reducing the illuminating aperture of the collimated beam to 25~mm diameter and recording spectra
of the etalon transmission at a series of positions across the etalon. The measuring positions
were spaced by steps of approximately 20~mm, producing 7 locations along the axis. The etalon was
then physically rotated $90^{\circ}$ to measure along the other axis. The effective gap at each
position was determined using equation~$\ref{intcond}$. The shape of the plates is expressed as
the change in the gap from that at the center of the plates. The measurements are plotted in
Figure~$\ref{flat}$.
For all three etalons, the effective gap decreases away from the
center. Moreover, this trend is generally wavelength dependent with the
plates somewhat flatter in the red than in the blue.

For the SG etalon, the plates exhibit roughly spherical curvature of approximately 15~nm along the
X-axis, but are flat along the Y-axis to within $\pm 1$~nm over most of the aperture, with a
turned-down edge of about 6~nm on one side. The weighted RMS deviation from the center (weighting
by the area of annular sectors) is about 8~nm at 650~nm ($\lambda/80$). For the MG etalon, the
effective shape along both the X and Y axes is more conical (i.e.\ the gap decreases roughly
linearly with radius), with a weighted RMS deviation of 12.2~nm ($\lambda/54$). Finally for the LG
etalon, the deviation along the X-axis is greater and more conical in the blue, smaller and more
spherical in the red. Along the Y-axis the plates are effectively flat and show no distinct shape.
Therefore the overall shape of the plates resembles that of a half cylinder. The weighted RMS
deviation at 650~nm is 13~nm ($\lambda/50$).

The effective shape measured here is not representative of the actual physical flatness of the
plates since interferograms of the plates taken by ICOS after polishing but before coating showed
flatness of $\lambda/100$ or better for all the plates. As discussed in Section~$\ref{coat}$, this
apparent curvature is likely due to an optical effect of the reflection coatings.

\subsubsection{Finesse and Resolving Power}\label{finres}
\begin{figure*}
  \includegraphics[angle=90,scale=0.65]{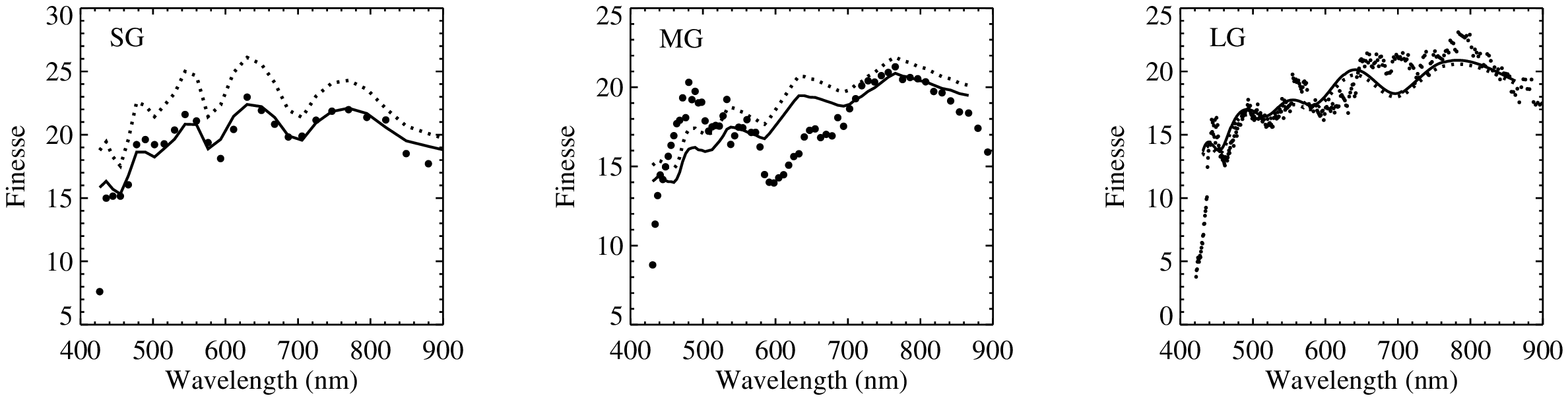}\\
  \caption{Finesse as a function of wavelength. Points represent
  the measured finesse. The curves are calculated from the reflectivity
  and measured (dotted) or fitted (solid) defect amplitude.}
\label{fin}
\end{figure*}

The significant deviations from flatness discussed in the previous section produce lower than
expected finesse and spectral resolution. If the only contribution to the total finesse comes from
$\mathcal{F}_{R}$, then using $R \sim 0.9$ (the average value from the coating curves in
Figure~$\ref{coating}$) and equation~$\ref{rfin}$ we would expect a finesse of $\mathcal{F} \sim
30$. But the measured finesse is less than the expected finesse because of the contribution of
defect finesse. The measured finesse for each of the three etalons is plotted as a function of
wavelength in Figure $\ref{fin}$.

For each etalon, we calculate a finesse curve using $\mathcal{F}_{R}$ from the measured
reflectivity, $\mathcal{F}_{D}$ for spherical curvature, and equation $\ref{tfin}$. We assume the
amplitude of the curvature defect is constant with wavelength. Then we either use the measured
weighted RMS value discussed in the previous section, producing the dotted curves in the figure,
or adjust the amplitude to optimize the fit, producing the solid curves.

For the SG etalon, the curve based on the measured defect amplitude of $\lambda/80$ has the
correct shape, but overestimates the actual finesse values. Adjusting $\delta t_{c}$ to be 10.8~nm
($\lambda/60$ at 650~nm) in equation~$\ref{dfin}$ yields the solid curve that fits the measured
finesse very well. This is our best definition of the effective flatness of the plates for this
etalon.

For the MG and LG etalons the shapes of the calculated curves do not agree as well with the
measured finesse distributions. As noted in the previous section, both the shape and the amplitude
of the surface defects change with wavelength, and these complications are not included in the
calculations. For both etalons the solid curve is calculated with a flatness of $\sim\lambda/49$
at 650~nm, which is very close the measured weighted RMS deviations of $\lambda/54$ and
$\lambda/50$.

\begin{figure*}
  \includegraphics[angle=90,scale=0.65]{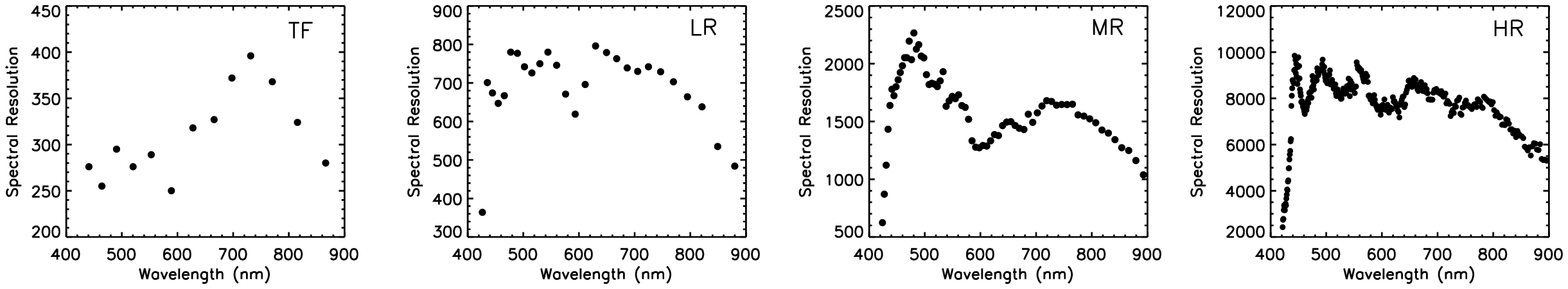}\\
  \caption{Plots of spectral resolution for the four spectral modes.
  The filled circles are the measured resolution for the individual orders.}\label{res}
\end{figure*}

The spectral resolution listed in Table~$\ref{main}$ is calculated using $\lambda/FWHM$. It is
related to the total finesse by equation~$\ref{rp}$. Thus the lower than expected total finesse
produces lower resolving power for all three etalons. Figure~$\ref{res}$ shows the spectral
resolution as a function of wavelength for all four resolution modes. The spectral resolution for
the LR, MR and HR modes is 780, 1500 and 8500 at 650~nm, respectively. Because these modes operate
at their largest possible gap setting, there is no way to increase the resolving power. The lowest
resolution TF mode has a resolving power of 350 at 650~nm. Since it is unwise to operate the
etalon with a smaller gap (to avoid bringing the plates into contact and damaging the plates
and/or their coatings), this is the lowest resolution available.
\subsubsection{Transmission}
\begin{figure}[t]
  \includegraphics[scale=0.5]{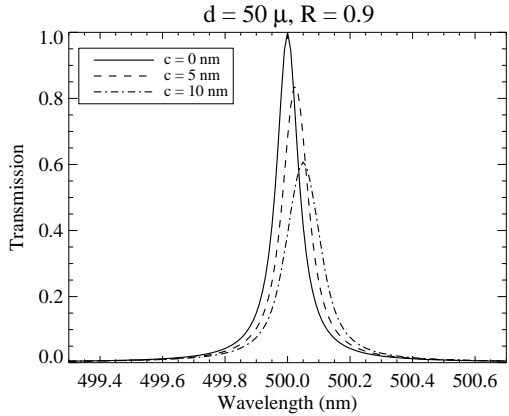}\\
  \caption{The effect of plate curvature on transmitted intensity}
  \label{transc}
\end{figure}

\begin{table}[t]
\centering
\begin{tabular}{ccccc}
\tableline \tableline
\\
$\lambda$ & R & $d$ & $I_{T}^{c\,\textrm{theory}}$ & $I_{T}^{\textrm{obs}}$\\
(nm) &  & ($\mu$m) & ($\%$) & ($\%$) \\
\tableline
440.0 & 0.892  & 4.18 & 68  &  66\\
455.2 & 0.886  & 4.34 & 72 & 68\\
505.0 & 0.912  & 4.29 & 64 & 57\\
555.0 & 0.916  & 4.44 & 65 & 58\\
579.0 & 0.888  & 4.63 & 79 & 72\\
641.5 & 0.914  & 5.13 & 72 &  63\\
665.1 & 0.896  & 6.33 & 80 & 63\\
705.2 & 0.880  & 7.42 & 86  &  75\\
735.2 & 0.892  & 7.37 & 84 & 77\\
783.5 & 0.896  & 7.05 & 85 & 78\\
815.2 & 0.885  & 6.94 & 88 & 81 \\
\tableline
\end{tabular}
\caption{This table makes comparison between the transmitted intensity calculated from equation
\ref{transcurv} (column 4) and the observed transmission values for the LR mode (column
5).}\label{trans}
\end{table}
The deviation of the plates from flatness also produces a decrement of the peak transmission. For
an ideal FP interferometer, the transmitted amplitude can be written as:
\begin{equation}
U_{t} = T\sum _{m = 0}^{\infty}
R^{m}\textrm{exp}\left(i\frac{4\pi}{\lambda}md\mu
\textrm{cos}\theta\right)
\end{equation}
where $T = 1 - R$ in absence of absorption. The product $U^{\dag} U$
gives the following transmitted intensity (Vaughan 1989):
\begin{equation}\label{transmission}
I_{t} = \frac{T^{2}}{1 - R^{2}}\left(1 + 2\sum_{m =
1}^{\infty}R^{m}\textrm{cos}(\frac{4\pi md}{\lambda})\right)
\end{equation}
assuming $\mu = 1$ and $\theta = 0^{\circ}$. This is basically the Airy function shown in
Figure~\ref{airy} (left panel). For a deviation from flatness in the form of spherical curvature,
the transmitted amplitude has the following form discussed by \cite{Mahapatra}:
\begin{equation}
U_{t}^{c} = T\sum_{m = 0}^{\infty}R^{m}\frac{\textrm{sin}(2\pi m
c/\lambda)}{2\pi m c/\lambda}\,\textrm{exp}\left[i\frac{2\pi
m}{\lambda}(2d + c)\right]
\end{equation}
where $c$ is the maximum deviation from a plane surface. Multiplying
this by its complex conjugate gives the following transmitted
intensity:
\begin{eqnarray}\label{transcurv}
\lefteqn{I_{t}^{c}  =  T^{2}\sum_{m = 0}^{\infty} R^{2m}\frac{\textrm{sin}(m\psi)}{m\psi}}
\nonumber\\
& &  \bigg\{1 + 2\sum_{k = 1}^{\infty} R^{k}\textrm{cos}(k\phi) \nonumber \\
& & \frac{m}{m + k}\frac{\textrm{sin}((m + k)\psi)}{\textrm{sin}(m\psi)}\bigg\}
\end{eqnarray}
where
\begin{equation}
\psi = \frac{2\pi c}{\lambda}\qquad \textrm{and} \qquad \phi =
\frac{4\pi d}{\lambda}(1 + \frac{c}{2d})
\end{equation}
equation~\ref{transcurv} reduces to equation~\ref{transmission} for $c=0$. Figure~\ref{transc}
shows the decrement in the transmitted intensity as $c$ increases.  Notice that the plate
curvature also broadens and shifts the transmission peak.

We measure the transmission from the ratio of two spectra: one with the etalon in the collimated
beam and the other with the etalon replaced by a 150~mm clear aperture. The measured transmitted
intensities for selected orders in the TF mode are listed in column~5 of Table~\ref{trans}. The
theoretical transmitted intensity is calculated using equation~\ref{transcurv} with $c = \delta
t_{c} =\lambda/80 = 8$~nm at 650~nm (taken from section \ref{pltflt} for the SG etalon) and is
listed in column~4 of Table~\ref{trans}. Although equation~\ref{transcurv} does not account fully
for the complicated effective curvature of the etalon plates, the calculated transmissions compare
reasonably well with those observed. The difference between the measured and calculated
transmitted intensity can be used to estimate the value of absorption, $A$, of the coatings. We
find $A \sim 0.5\%$, a typical value for multi-layer dielectric high-reflection coatings.

\subsubsection{Wavelength Calibration}
The etalon is scanned in wavelength by changing the controller $Z$ offset value, which causes the
gap size, $d$, to change. In the lab, we take spectra of the etalon's transmission at a range of
$Z$ settings to calibrate the $Z$ -- wavelength relation. (At the telescope, a different
calibration procedure is used, where the etalon is tuned to the wavelengths of several emission
lines from spectral lamps and the resulting ring patterns are measured.) We find that a low-order
polynomial provides an excellent fit to the wavelength calibration:
\begin{equation}
\lambda = A + BZ' + CZ'^{2} + DZ'^{3},\quad\quad
\end{equation}
\noindent where
\begin{equation}
\quad Z' = \frac{Z}{1000}
\end{equation}

The scanning range of the $Z$ offset ($-2048$ to $+2047$) is sufficient to cover several FSRs of
an etalon. For the MG and LG etalons we use this to limit the number of orders calibrated for
operational simplicity while retaining the ability to tune to any desired wavelength within the
available bandpass (430 to 900~nm).  We calibrate 12 orders of MG and 54 of LG.  Over the full $Z$
range a cubic polynomial fits best, with residuals of less than 1/40 of the FWHM. We operate the
SG etalon in two modes: TF and LR, near the minimum and maximum gap, respectively.  To maintain
these extremes, we limit the $Z$ range to that required to scan a single order ($-1.250$ to
$-0.750$ and $+1.260$ to $+2.000$ respectively) and calibrate every order: 13 in TF and 27 in LR.
Over these restricted scanning ranges a quadratic polynomial provides an excellent fit, with
residuals less than 1/50 of the FWHM.

The $A$ coefficient is the wavelength `zero point' and can vary with time, temperature, humidity
and other environmental factors. Its stability will be studied extensively in the telescope
environment during commissioning. We expect that a single zero point calibration will be required
on hourly time scales while observing. Our experience suggests that the $B$, $C$ and $D$
coefficients will be very stable, requiring no more than weekly or monthly recalibration.

\subsubsection{Stability}
\begin{figure*}
  \includegraphics[angle=90,scale=0.65]{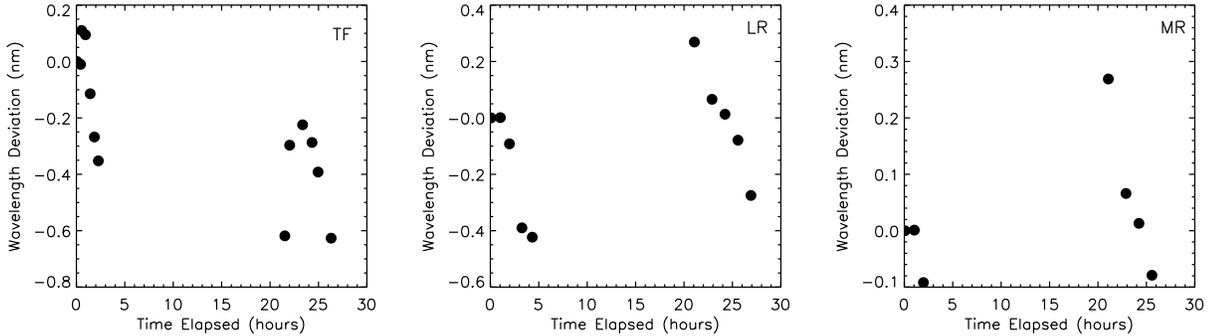}
  \caption{Temporal stability of the etalon zero point for TF, LR
   and MR modes}\label{stab}
\end{figure*}
The wavelength stability, which is essentially the stability of the zero point $A$, of the etalons
was measured by taking a series of spectra over a period of more than 24 hours. During this period
the etalon and the controller were undisturbed mechanically and electrically and the environmental
conditions (like temperature) were held approximately constant. The measurements, plotted in
Figure $\ref{stab}$, were carried out for the TF, LR and MR modes only. The wavelength drift for
these modes were smooth and typically $\sim 0.1$~nm per hour. Over the entire period of
measurement the RMS wavelength fluctuations in the TF, LR and MR modes were 0.24~nm, 0.22~nm, and
0.071~nm, respectively. Note an abrupt jump of 0.3~nm in the zero point of LR mode between hour 2
and 3; there was a power failure in the lab which caused the controller to reset.

Additionally, the RMS FWHM fluctuations over the entire period were 0.075~nm in TF, 0.033~nm in LR
and 0.021~nm in MR modes. For the MR mode, we also measured the parallelism stability over a 7-day
period. The RMS variations in the $X$ and $Y$ offsets over this period were 4.1 units and 2.4
units, respectively. These drifts are insignificant because the uncertainty in determining the
optimum parallelism is at least 5 units. The wavelength and parallelism stability may change with
variable environmental conditions and hence will be tested at the telescope during commissioning.

\section{Discussion: Effect of Phase Dispersion on Multi-layer Broadband Coatings}\label{coat}

Our laboratory testing uncovered three unexpected results: wavelength-dependent parallelism
changes, wavelength-dependent gap changes, and significant apparent plate curvature. We believe
that all of these phenomena arise from effects in the multi-layer broadband reflection coatings of
the etalons.

\subsection{Broadband Multi-Layer Coating Design}
Although the exact coating structure of our etalons is proprietary information and we are unable
to present it here, we have been given access to some of the design details to simulate the
properties of the real coating using multi-layer design software. The coatings comprise 16 layers
of Zinc Sulphide and Cryolite and are deposited without heating the substrates. The total
thickness of the coating is 1331~nm.

The evolution of coating design from the simple quarter-wave layer AR coating to the fully
optimized broadband multi-layer has been facilitated by the advent of proprietary and commercial
coating design algorithms. Nevertheless, coating design and deposition is still an art as well as
a science, requiring experience and an intimate knowledge of the performance and properties of the
coating vendor's deposition chamber.

\begin{figure}[t]
  \includegraphics[scale=0.65]{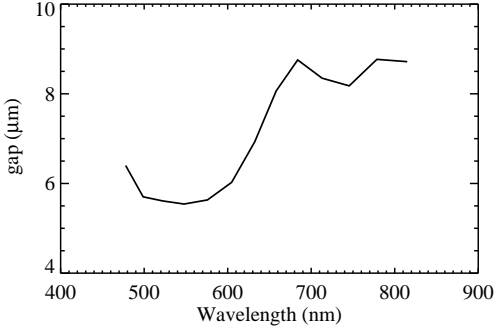}
  \caption{Predicted effective gap variation with wavelength, based on our coating design and
    a 5~$\mu$m air gap.}
  \label{coatgap}
\end{figure}

An early scheme for extending the spectral range of coatings was to combine two quarter-wave
stacks on top of each other, with the two stacks optimized for different wavelength bands. To
minimize scattering effects, which are usually more severe at shorter wavelengths, the upper stack
was designed to reflect the shorter wavelengths, while the lower stack reflected the longer
wavelengths. Thus longer wavelengths penetrated deeper into the coatings, giving rise to a larger
effective gap at these wavelengths. While such design techniques are no longer necessary, it
appears that the fully optimized coating covering our spectral range of 430--860~nm shows relics
of this approach. Our measurements indicate that the transition occurs in the 600 to 720~nm
wavelength region with the additional path as large as that due to the optical thickness of
2.6~$\mu$m of coating material. We have simulated this effect for our coating designs with a
5~$\mu$m air gap, producing the curve in Figure $\ref{coatgap}$.  Comparison with Figure
$\ref{gapfsr}$ shows an excellent agreement of this prediction with our measurements of the
effective gap. \cite{ryder} observe similar effects for etalons coated for the near infrared.

\subsection{Phase-Dispersion}
\begin{figure*}
  \includegraphics[scale=0.85]{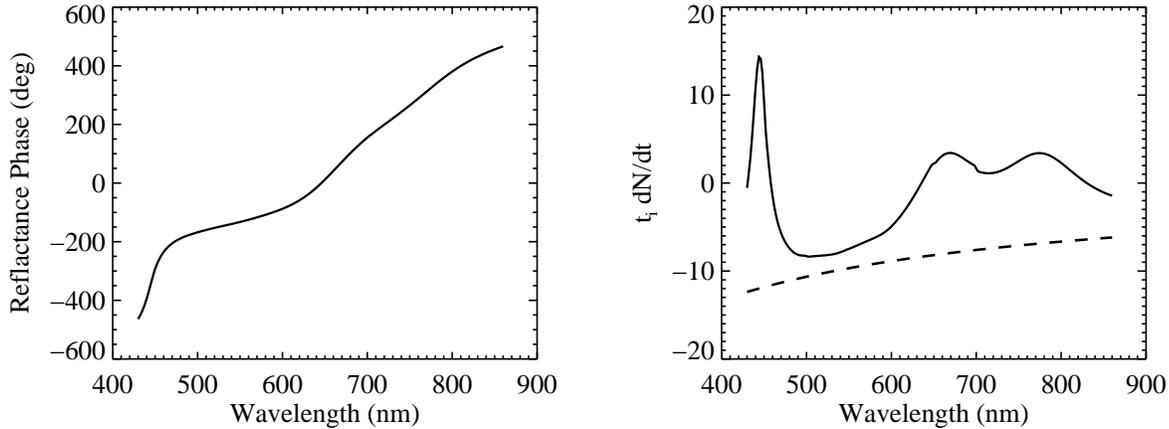}
  \caption{Left: calculated phase shift for our coating design.  Right: Amplification effect
  for these coatings; solid curve is full Equation $\ref{phase}$, dotted curve is the geometrical
  term alone.}
  \label{phasedisp}
\end{figure*}

In multi-layer dielectric reflectors, a wavelength-dependent phase shift ($\beta$) occurs upon
reflection. \cite{ramsay} investigated the effect of phase-shift dispersion
($\partial\beta/\partial\lambda$) in multi-layer dielectric reflectors on the apparent shape of
the reflecting surface. They report an experiment where a plate with a physical surface of
$\lambda/60$ concave, when coated with a standard multi-layer reflector, showed an effective
surface shape of $\lambda/60$ concave at wavelength 546~nm, flat at 588~nm, and $\lambda/10$
convex at 644~nm! The phase-shift dispersion can amplify the effects of slight irregularities in
the coating thickness or the substrate surface to produce a significant, and wavelength dependent,
change in the apparent shape of the surface. \cite{ramsay} present a result from \cite{giacomo}:
\begin{equation}\label{phase}
t_{i}\frac{\partial N}{\partial t_{i}} = -\left(\frac{4t}{\lambda} -
\frac{\lambda\partial\beta}{\pi\partial\lambda}\right)
\end{equation}
where $t$ is the geometrical thickness of a multi-layer stack, $i$ refers to the different layers
and $N$ is the interference order related to gap by equation \ref{intcond} of this paper. The term
$\frac{4t}{\lambda}$ is the geometric effect, and is independent of the phase dispersion in the
coatings.
Figure $\ref{phasedisp}$ shows the expected phase-shift curve as a function of wavelength for our
coating design.
Using this phase dispersion and $t = 1331$~nm, Equation \ref{phase} is plotted in the right
panel of Figure \ref{phasedisp}; for comparison, we also show the geometrical term alone.

It is apparent from Equation $\ref{phase}$ that the phase dispersion term can reduce or cancel the
geometrical term, decreasing the sensitivity to surface and coating imperfections. \cite{ciddor}
discusses attempts to design coatings with phase dispersions that cancel the geometric effect, and
thus eliminate or significantly reduce the apparent curvature of the plates. He comments that it
is very difficult (if not impossible) to simultaneously maintain a uniform high reflectivity over
a broad bandwidth while producing the desired phase dispersion cancelation. Figures
\ref{phasedisp} and \ref{coating} suggest that our coatings have partially achieved this goal. The
sensitivity to imperfections is reduced throughout the bandpass except for very small range
between 442~nm and 448~nm where the amplification is only slightly worse than the geometrical term
alone (but of the opposite sign).  Throughout the red half of the bandpass the sensitivity is
reduced by about a factor of 3.  These results are only approximate, since we do not have the
as-built details of the coatings, but we believe that they are indicative.

The manufacturing tolerance in the individual layer thickness of coatings available commercially
is often quoted as $< 2\%$, but this is a conservative figure and today's coatings are undoubtedly
better. \cite{Netterfield} show that with great care, modern coating techniques can achieve
$\partial t /t \sim 0.001$. Our measured surface curvatures of RMS amplitude 8 -- 13~nm are
consistent with the amplification factors of -8 to +3 (from Figure $\ref{phasedisp}$) and coating
thickness variations of $\sim 0.7\%$. To keep the phase dispersion amplified distortions to be
less than $\lambda/100$ (the physical flatness of our plates) would require coating thickness
variations of less than 0.3\% - a demanding but achievable goal.

The change of parallelism with wavelength may also arise from a similar effect. This change occurs
in precisely the same wavelength region as the apparent gap change, and so is presumably related
to the same sub-structure in the coatings. The similarity of the shape of the parallelism curves
for all three etalons, shown in Figure~\ref{para}, suggests that this is a systematic, not random
effect. Thus one might suspect that there is a systematic tip between the upper and lower layers
of the coating. The measured amplitude of the parallelism changes is $-78 \pm 20$~nm in X and $+59
\pm 7$~nm in Y. The coating irregularities deduced in the previous paragraph were of order 9~nm,
and the effect of Equation~\ref{phase} could amplify these by factors $\sim10$, so it is possible
that this could produce the amplitude of the effect. The shape of the amplification curve in
Figure $\ref{phasedisp}$ (constant at short and long wavelengths and changing rapidly between 600
and 700~nm) is suggestive of the shape of the parallelism curves; however we do not observe any
significant parallelism change around the steep peak at 440~nm of the amplification curve. The
plates are marked and their orientation is preserved through the polishing, coating, and assembly
processes, so the similarity of the shapes of the parallelism curves for all three etalons could
well arise from reproducible irregularities inherent in the coating process. Thus we speculate
that the phase dispersion amplification of a relatively small wedge in the coating layers may be
the origin of the parallelism changes.

We conclude that it is critical to take into account the effects of phase dispersion when
designing broadband coatings for high-finesse FP systems.  In our system, the unanticipated
effects of the coatings are the main factor that limit the performance of the system. While some
specialized FP systems require no relative phase shift over a particular bandwidth (e.g.
\cite{lem}), it seems that for general use tailoring the phase dispersion to minimize the
amplification effects of equation~\ref{phase} is highly desirable. Although it is unlikely that
the competing demands of uniform high reflectance over a broad band and a phase dispersion that
cancels the coating effects over the bandwidth can be exactly satisfied, it should be possible to
find designs that provide a reasonable compromise.

\section{Conclusions}\label{conc}
In this paper we have presented the design of the SALT FP system, and the laboratory measured
characteristics of the as-built system. The system provides spectroscopic imaging at any desired
wavelength from 430 to 860~nm in four resolution modes from 300 to 9000. We discovered that the
plate parallelism and gap and are wavelength dependent with maximum variation between wavelengths
600 to 720~nm; there is more apparent plate curvature than expected, leading to lower resolution,
finesse, and throughput. We believe that all these effects arise from the design of the broadband
reflectance coatings of the etalons. It is unfortunate that the $H\alpha$ line falls within the
wavelength region of maximum gap and parallelism variation. This will require more care in
operating the etalons for many scientific programs, but will not seriously compromise the
resulting data. The design goals for the system were based on surveys of the SALT user community,
and represent typical but not mandatory requirements for a broad range of scientific
investigations. The as-built etalons approach but do not fully meet these goals. The performance
of the system is certainly sufficient to satisfy the needs of a very wide variety of programs for
the SALT observers, and users will undoubtedly design future investigations to exploit the
capabilities of the system.

The SALT FP system is presently in its performance verification phase at the telescope and the
detailed on-sky performance of this system will be presented in a subsequent paper.

\begin{acknowledgements}

The vision and leadership of Robert Stobie provided the concept of the SALT project and brought it
into being. Without his keen scientific insight and talent for management none of the work
described in this paper would have been undertaken. Ken Nordsieck is the leader of the RSS
instrument team and has provided valuable advice and support for the development of the FP
subsystem. The assistance of Eric Burgh, as instrument scientist of RSS, has been crucial for many
aspects of the work described here. David Buckley, the SALT Project Scientist and the Chair of the
SALT Science Working Group, oversaw the specification of the scientific requirements for the
instrument and managed its development. The SALT construction team in South Africa and the RSS
instrument team at the University of Wisconsin -- Madison have contributed in many ways to this
project. Terry Dines of ICOS is responsible for producing the state-of-the-art plates that are the
heart of the RSS FP. David Marcus of Custom Scientific, Inc. provided much useful advice in the
design of the filter set. Funding for the RSS FP system was provided by an anonymous donor and by
the Dean of the Faculty of Arts and Sciences of Rutgers University.
\end{acknowledgements}

\newpage


\begin{thebibliography}{}        
\bibitem[Swat et al.(2003)]{arek} Swat, Arek, O'Donoghue, Darragh, Swiegers, Jian, Nel, Leon, Buckley,
David A. H. 2003, Proceedings of the SPIE, 4837, 564
\bibitem[Atherton et al.(1981)]{atherton} Atherton, P.D., Reay, N.K., Ring, J. \& Hicks, T.R. \
1981, Optical Engineering, 806, 20
\bibitem[Burgh et al.(2003)]{eric} Burgh, Eric B., Nordsieck, Kenneth H., Kobulnicky, Henry
A.,Williams, Ted B., O'Donoghue, Darragh, Smith, Michael P., Percival, Jeffrey W. 2003,
Proceedings of the SPIE, 4841, 1463
\bibitem[Cecil (2000)]{aries} Cecil, G. N. 2000, Proceedings of the SPIE, 400, 83
\bibitem[Ciddor(1968)]{ciddor} Ciddor, P.E. 1968, Applied Optics, 7, 2328
\bibitem[Fowles(1989)]{fowles} Fowles, G.R. 1989, Introduction to
Modern Optics, (Dover Publications, INC., NY)
\bibitem[Giacomo (1958)]{giacomo} Giacomo, P. 1958, Le Journal De Physique Et Le Radium, 19, 307
\bibitem[Kobulnicky et al.(2003)]{Kob} Kobulnicky, Henry A., Nordsieck, Kenneth H., Burgh, Eric B., Smith,
Michael P., Percival, Jeffrey W., Williams, Ted B., O'Donoghue, Darragh 2003, Proceedings of the
SPIE, 4841, 1634
\bibitem[Lemarquis \& Pelletier (1996)]{lem} Lemarquis, F. \& Pelletier, E. 1996, Applied Optics, 35, 4987
\bibitem[Mahapatra \& Mattoo(1986)]{Mahapatra} Mahapatra, D.P., \&
Mattoo S.K. 1986, Applied Optics, 25, 1646
\bibitem[Netterfield et al. (1980)]{Netterfield} Netterfield, R. P., Schaeffer, R. C. \& Sainty,
W. G. 1980, Applied Optics, 19, 3010
\bibitem[Nordsieck et al.(2003)]{Ken} Nordsieck, Kenneth H., Jaehnig, Kurt P., Burgh, Eric B., Kobulnicky,
Henry A., Percival, Jeffrey W., Smith, Michael P. 2003, Proceedings of the SPIE, 4843, 170
\bibitem[Ramsay \& Ciddor (1967)]{ramsay} Ramsay, J.V. \& Ciddor,
P.E. 1967, Applied Optics, 6, 2003
\bibitem[Schommer et al. (1993)]{rfp} Schommer, R.A., Bothun, G.D., Williams, T.B., \& Mould, J.R. 1993,
AJ, 105, 97
\bibitem[Smith et al.(2006)]{mike} Smith, Michael P., Nordsieck, Kenneth H., Burgh, Eric
B.,Percival, Jeffrey W., Williams, T. B., O'Donohue, Darragh, O'Connor, James, Schier, J. Alan
2006, Proceedings of the SPIE, 6269, 62692A
\bibitem[Stobie, Meiring \& Buckley(2000)]{stobie} Stobie, Robert, Meiring, Jacobus, Buckley, David
A. 2000, Proceedings of the SPIE, 4003, 355
\bibitem[Vaughan (1989)]{vaughan} Vaughan, J. M. 1989, The Fabry-P\'{e}rot Interferometer,
(Adam Hilger, Philadelphia)
\end{thebibliography}
\end{document}